\documentclass[prx,twocolumn,superscriptaddress,12point,longbibliography]{revtex4-2}
\usepackage{amsmath,amssymb,bm}
\usepackage{wasysym}
\usepackage{graphicx}
\usepackage{epstopdf}
\usepackage{latexsym}
\usepackage{subfigure}
\usepackage[usenames, dvipsnames]{color}
\usepackage[usenames, dvipsnames]{xcolor}
\usepackage{natbib}
\usepackage{braket}
\usepackage{float}
\usepackage[normalem]{ulem}
\usepackage{comment}
\usepackage{mathtools}
\usepackage{array}
\usepackage{tabu}
\usepackage{multirow}
\usepackage{chemformula}
\usepackage{svg}
\usepackage[T1]{fontenc}
\usepackage[	
citecolor=ForestGreen,
colorlinks, 
linkcolor=violet,
urlcolor=black,
]{hyperref}
\newcommand\redsout{\bgroup\markoverwith{\textcolor{red}{\rule[0.5ex]{2pt}{0.4pt}}}\ULon}
\newcommand\bluesout{\bgroup\markoverwith{\textcolor{black}{\rule[0.5ex]{2pt}{0.4pt}}}\ULon}

\newcommand{\SPhide}[1]{{}}

\def\acz2{ac-$\mathbb{Z}_2$}

\begin{document}

\title{Toric code made subsystem: a 
framework for topological subsystem codes 
using anticommuting quantum spin liquids}
\author{Vaibhav Sharma}
\affiliation{Smalley-Curl Institute, Rice University, Houston, TX 77005}
\affiliation{Department of Physics and Astronomy, Rice University, Houston, TX 77005}
\email{vaibhavsharma@rice.edu}
\author{Sumiran Pujari}
\affiliation{Department of Physics, Indian Institute of
Technology Bombay, Mumbai, MH 400076, India}
\email{sumiran.pujari@iitb.ac.in}

\begin{abstract}
We introduce a framework of constructing topological subsystem codes based on the class of  anticommuting~quantum spin liquids described in \href{https://journals.aps.org/prb/abstract/10.1103/dhh4-8wkj}{[Phys. Rev. B 113, 064402 (2026)]}. A canonical model from this class can be considered as a spatial modification of the toric code that voids its stabilizer code property. Rather, these models contain an extensive set of anticommuting local conserved operators that lead to an extensive ground state degeneracy. 
This degeneracy forms the basis of the subsystem degrees of freedom in the associated quantum error correcting code. 
The code inherits the many-body topological order of the quantum spin liquid, making it a topological subsystem code. 
We present two concrete and detailed examples for constructing these codes on a square lattice and a kagome lattice geometry, requiring weight-4 and weight-3 local check operator measurements respectively. 
In contrast to other subsystem codes, a unique property of these codes is the presence of an extensive number of local gauge qubits that are left undisturbed by the check operators apart from the logical qubits.
Our construction provides a template for generating this new category of topological subsystem codes on different lattice or graph geometries, suitable for implementation on various quantum hardware platforms.
\end{abstract}

\maketitle

\section{Introduction}
\label{sec:intro}

The fields of quantum condensed matter and quantum information processing have had many fruitful exchanges for over two decades now. 
Fault tolerance and error correction in quantum computing is a noteworthy case of such exchange and also the subject of this work.
A canonical example is the toric code or the surface code that originated with a goal towards fault tolerant quantum computation while informing greatly on the subject of many-body topological order (in particular $\mathbb{Z}_2$ lattice gauge theories)~\cite{Kitaev_2003}. 
Furthermore, the possibility of non-local Majorana zero modes in quantum matter serves as another example where knowledge created in condensed matter physics has been exploited to design quantum information processing platforms~\cite{topoqecreview}.
More generally, many-body topological order has become one of the foundations for topological quantum computation and quantum error correction (QEC).

\vspace{0.5cm}
The theory and formalism of QEC is by now well-developed over the past two to three decades~\cite{qecreview}.
Several QEC codes and protocols have already been implemented in experiments~\cite{qecexp1,qecexp2,qecexp3,qecexp4,qecexp5,qecexp6,qecexp7,qecexp8}, demonstrating increased accuracy of quantum operations.
The basic idea of any QEC code or protocol is as follows: 
1) An encoding scheme whereby the logical qubit is encoded or represented by an appropriately chosen configuration of underlying physical qubits. 
2) An error detection scheme through the design of appropriate ``syndrome'' measurements or operators that operate on the physical qubits and return a yes/no answer to indicate the presence or absence of an error in the logical qubit. 
And finally, 3) an error correction scheme that gives the appropriate operation or sequence of operations to correct the detected error.
An integral part in the above schema is the choice of the underlying physical qubit configuration in terms of a hardware implementation . 
This often involves identifying appropriate geometry or arrangement of the physical qubits in physical space as a practical matter. 
This is apart from general and relatively more abstract QEC code design steps such as which particular states or set of states would represent the logical $0$ and $1$, and/or the choice of check/syndrome and the logical operators.

In this work, we will concern ourselves with the QEC properties of a class of quantum spin liquid model Hamiltonians that was recently proposed in the quantum matter literature and termed as anticommuting $\mathbb{Z}_2$ quantum spin liquids (ac-$\mathbb{Z}_2$ QSLs)~\cite{anticommutingqsl,GSentropy_preprint}.
Such model Hamiltonians apart from being spin liquids can also host many-body topological order.
As is well-known, many-body topological order is one of the natural basis for fault tolerance going back to the toric code proposal~\cite{Kitaev_2003,qecreview1}.
However, the ac-$\mathbb{Z}_2$ QSLs apart from potentially hosting topological order \emph{always} possess an exponential spectral degeneracy that leads to the spin liquidity, i.e. these spin liquids are gapless with an extensive residual entropy.
Fault tolerance and QEC properties of the toric code relies on the presence of the topologically protected gap to excitations.
Hence, \textit{a priori}, it is not clear if these gapless QSLs are viable for QEC or not.
Thus the central question that we will be addressing in this work is the following: Do any ac-$\mathbb{Z}_2$ QSLs possess QEC properties when viewed as a code?

An example of a ac-$\mathbb{Z}_2$ QSL Hamiltonian is given by 
\begin{equation}
    H =\; J_x \sum_{\boxed{x}} \left( \prod_{i \in\; \boxed{x}} 
    \sigma^x_i \right)
    +
    J_z \sum_{\boxed{z}} \left( \prod_{j \in\; \boxed{z}} 
    \sigma^z_j \right)
    \label{eq:2dmodelapintro}
\end{equation}
where qubits are laid out on the vertices of a square lattice 
(see Fig.~\ref{fig:latticecheckops} for a guide).
The operators $\sigma^\mu_i$ are Pauli operators on site $i$ and $\boxed{\mu}$ indicates square plaquettes of ``$\mu^{\text{th}}$'' type.
The toric code Hamiltonian in comparison would have the qubits located rather at the bonds of the lattice which lends it a commuting projector Hamiltonian structure, i.e. a stabilizer code in terms of QEC.
Despite the apparent similarity of Eq.~\ref{eq:2dmodelapintro} to the toric code, the structure is not of the commuting projector type anymore due to the corner-sharing nature of the ``nearest-neighbour plaquette'' Hamiltonian terms.
This kind of spatial arrangement is a key difference that distinguishes the class of ac-$\mathbb{Z}_2$ qubit Hamiltonians and will govern their QEC properties.
Apart from multi-spin couplings such as in Eq.~\ref{eq:2dmodelapintro} above, ac-$\mathbb{Z}_2$ QSLs with two-spin couplings are also possible but have not been shown to possess topological order~\cite{anticommutingqsl}.

We now state the answer to the question posed above: ac-$\mathbb{Z}_2$ QSL Hamiltonians that possess many-body topological order are realizations of topological subsystem codes when viewed as QEC codes.
These codes are an amalgamation of two separate notions, i.e. that of a topological code such as the toric code~\cite{Kitaev_2003} and that of a subsystem code~\cite{Poulin_2005,Kribs_Laflamme_Poulin_2005,Kribs_etal_2006} such as the Bacon-Shor code~\cite{baconshor}.
The logical qubits are encoded non-locally among the physical qubits that is furthermore topological in nature. 
This is very much in the spirit of the toric or surface code, and the topological encoding of the logical qubits will endow them with fault tolerance from environmental errors or noise that are locally restricted.
The possibility of topological encoding is essentially inherited from the many-body topological order of the chosen ac-$\mathbb{Z}_2$ QSL construction.
We will see how this is quite natural for qubit models like Eq.~\ref{eq:2dmodelapintro}. 
The subsystem nature of these codes, it turns out, is linked to the massive spectral degeneracy related to the spin liquidity of the QSL Hamiltonians. 
This is guaranteed from anticommuting local conserved operators that are present by construction in these models~\cite{GSentropy_preprint,anticommutingqsl}. 
We can refer to codes constructed within this framework as anticommuting charge (ACC) codes, where ``charge'' denotes the local conserved operators.

The subsystem property of these codes can offer some unique advantages~\cite{subsystem1}. 
In subsystem codes, there are several logical qubits but some of them do not encode any information. 
These are referred to as redundant gauge qubits. 
This allows the gauge qubits to absorb some errors without affecting the encoded information and simplify the error correction procedure~\cite{subsystem1}. 
For example, the Bacon-Shor subsystem code requires only two-qubit measurements for syndrome readout while performing error correction~\cite{baconshor}, as opposed to four-qubit measurements needed in the toric code. 
The Bacon-Shor code has already been implemented in experiments with trapped-ion qubits~\cite{expbaconshor1,expbaconshor2}. 
Furthermore, proven upper bounds show that 2D subsystem codes with local gauge qubits can encode more logical qubits than analogous stabilizer codes~\cite{bravyibound}. 

Topological subsystem codes, such as the ones constructed in this work can thus combine fruitful properties of both topological stabilizer codes and subsystem codes. 
They were first proposed by Bombin~\cite{Bombin_2010,Bombin_2015} on trivalent lattices, followed by the construction by Bravyi \emph{et al} on a square lattice called the subsystem surface code~\cite{toposubsys}. Given the relation of Bravyi \emph{et al}'s subsystem surface code to the surface code, the titular phrase ``toric code made subsystem'' refers to the similitude of our particular code constructions (cf. Eq.~\ref{eq:2dmodelapintro}) to the toric code, except for the innocuous looking change from the bond-sharing property to the site-sharing property.
As already remarked above, this is going to be crucial for the subsystem properties of our code constructions.
Additionally, these prior proposals for topological subsystem codes appear somewhat as one-off constructions with Bombin's scheme allowing for some variety despite restricting to trivalent graph connectivity. 
Our code constructions rely primarily on the kind of corner-sharing property discussed above when it comes to the physical layout of the constituent qubits.
This lends more flexibility in code construction and reduces qubit count when compared to, for instance, Bravyi \emph{et al}'s subsystem surface code on the square lattice~\cite{toposubsys}. 
Furthermore in our code, there are by construction an extensive number of gauge qubits (the local anticommuting charges) that are left undisturbed when measuring the code stabilizers through local check operators. 
This is a distinct feature that is generically absent in standard subsystem codes.

In our work, we will argue thus that ac-$\mathbb{Z}_2$ QSL Hamiltonians provide a natural template for realizing a new class of topological subsystem codes as motivated in the preceding discussion. 
The ingredients in our template can be used to find topological subsystem codes in a variety of settings that could potentially enhance encoding rates or error threshold rates. 
This may also be helpful for implementation on quantum hardware setups that might have different preferred connectivities and geometrical layouts that are not amenable for other rigid code constructions.
We discuss this some more in the context of existing hardware in the final section (Sec.~\ref{sec:conclusion}).
Our paper is organized as follows: in Sec.~\ref{sec:review}, we briefly review subsystem codes and when are they topological.
In Sec.~\ref{sec:code_construction}, we lay down the general principles that go behind the construction of topological subsystem codes from the ac-$\mathbb{Z}_2$ QSL Hamiltonians.
In Sec.~\ref{sec:weight4},~\ref{sec:weight3}, we discuss in detail two specific ACC codes that are weight-4 and weight-3 on a square and kagome lattice respectively. Eq.~\ref{eq:2dmodelapintro} inspires the weight-4 code to be discussed in Sec.~\ref{sec:weight4}.
Finally, we conclude with a brief discussion and outlook in Sec.~\ref{sec:conclusion}.

\section{Brief review of topological subsystem error-correction codes}
\label{sec:review}

In this section, we first recap subsystem quantum error correction codes and then briefly discuss the concept of topological subsystem codes. This would introduce the terms and language that we use throughout the paper. For a review of stabilizer subspace codes such as the surface code, see Ref.~\cite{qecreview}. The recent tutorial review~\cite{qecreview1} gives a detailed, pedagogical introduction to these topics.

\subsection{Subsystem error correcting codes}

A subsystem error correcting code encodes logical information in a \textit{subsystem} within the physical qubit system as opposed to a particular \textit{subspace} within the entire Hilbert space. For example, consider a Hilbert space $\mathcal{H}$ which can be decomposed as, $\mathcal{H} = \mathcal{C} \otimes\mathcal{D}$. We encode logical information in the subsystem $\mathcal{C}$ while ignoring the state of the qubits in $\mathcal{D}$. The degrees of freedom in $\mathcal{D}$ can vary freely and are termed as \textit{gauge} degrees of freedom. Any operations on $\mathcal{D}$ act trivially on the code subsystem $\mathcal{C}$. 
The standard presentation is sometimes given as $\mathcal{H} = \left(\mathcal{C} \otimes\mathcal{D} \right) \oplus \mathcal{E}$ where $\mathcal{E}$ represents the ``rest of the universe'' or the surrounding environment. For the subsequent discussions, this detail is not going to be relevant.

Errors in the encoded information are diagnosed by measuring code stabilizer operators whose syndrome detects the location and type of the error. 
In a subsystem code, the code stabilizers can be non-local operators but they are generated by  local operators that act trivially on the code subsystem $\mathcal{C}$. 
One can think of it as the non-local code stabilizers being measured through multiple local operators. 
These local operators are referred to as \textit{check} operators and measuring them lets us measure the code stabilizers for error detection. 
An error correction procedure is then applied such that the effect of the error followed by the error correction acts as identity on the code subsystem $\mathcal{C}$. 
This procedure may cause some non-trivial operation on the gauge subsystem $\mathcal{D}$. 
This has no effect on the logical qubit encoding because it occurs in the redundant subsystem $\mathcal{D}$. 
This redundancy can also be interpreted as the logical code space being massively degenerate. 
This is in contrast to a subspace code such as the toric code where a logical qubit's code space is unique and the error correction procedure restores us back to the same state within the entire Hilbert space.

A paradigmatic example of such a subsystem code is the Bacon-Shor code defined on an $L \times L$ square lattice~\cite{baconshor}. 
It is referred to as an $[n,k,d]$ code where $n=L^2$ is the number of physical qubits, $k=1$ denotes the number of logical qubits and $d=L$ is the code distance. 
The code stabilizers and logical qubit operators are non-local weight-$L$ operators. 
The check operators are local two-qubit Pauli XX and ZZ operators. 
They do not commute with each other when they have a single common qubit, forming a non-Abelian group. 
They are used to measure the code stabilizers for error syndrome detection. 
Due to the lower weight check operators, subsystem codes are more hardware-friendly and have a natural protection against any errors in the gauge subsystem. 
They are also useful in circumventing certain theorems limiting the number of logical qubits in stabilizer subspace codes with spatially local stabilizers~\cite{bravyibound}. 
The subsystem degree of freedom further lends flexibility in detecting and correcting errors~\cite{qecreview1}.

\subsection{Topological subsystem codes}

For a subsystem code to be a topological subsystem code, it must possess the following topological aspects: 
(i) the number of encoded qubits must be sensitive to the topology of the global lattice or graph connectivity in view of fault tolerance, and as a corollary, 
(ii) the code should be flexible such that it can be defined on different lattices or graphs~\cite{Bombin_2010}. 
The Bacon-Shor code is a subsystem code but it is not a topological code as it is rigid and only encodes one logical qubit regardless of whether, for example, the lattice has periodic or open boundary conditions which changes the lattice topology. 
In contrast, the toric code is a $[L^2,k,L]$ topological stabilizer code on an $L \times L$ square lattice as the number of encoded qubits, $k$ is directly determined by how the lattice wraps around the boundary: 
$k=2$ when defined on a torus and $k=1$ when one of the lattice directions does not have periodic boundary conditions. 
The case of $k>2$ is also possible by ``punching additional holes'' in the lattice~\cite{qecreview,toposubsys}.

A topological subsystem code can be considered as a combination of a topological stabilizer code such as the toric code and a subsystem code.
Topological subsystem codes were first introduced in Ref.~\cite{Bombin_2010} with two-qubit check operators on tripartite lattices. 
They have also been shown as generalizations of Kitaev honeycomb code on 3-valent hypergraphs~\cite{SBT11,Sarvepalli_Brown_pra2012} and further extended to square lattice geometry as the subsystem surface code~\cite{toposubsys}. 
In this work, we provide an intuitive scheme to construct topological subsystem codes on different lattice or graph geometries that neither need be 3-valent nor square, inspired by the structure of anticommuting quantum spin liquids. 
This allows them to be easily interpreted and may be implemented on experimental setups featuring different connectivity geometries. 

\section{Constructing the code from the anticommuting spin-liquid Hamiltonian}
\label{sec:code_construction}

Here we describe how the model Hamiltonians of anticommuting quantum spin liquid states include the necessary ingredients to construct a topological subsystem code. 
We refer to the class of anticommuting spin liquid Hamiltonians introduced in Ref.~\cite{GSentropy_preprint,anticommutingqsl} and focus on 2D lattices throughout. 
The defining characteristic of these spin liquids is the presence of an extensive number of local conserved operators that do not mutually commute with each other. 
This leads to an extensive ground state degeneracy. 
The ground state can be ensured to have no local order but only non-local order, making it a topological quantum spin liquid. 
The presence of an extensive degeneracy with topological order is suggestive of the gauge degrees of freedom in a subsystem code that we intend to exploit here. 
This will form the core basis of our construction of topological subsystem codes from anticommuting quantum spin liquids. 

Formally, we first find the smallest weight operators that commute with all the local anticommuting conserved operators in the spin liquid model. 
These are the code's check operators. 
This choice ensures that the code possesses the same topological properties as the quantum spin liquid model from which it is inspired. 
The check operators need not mutually commute among themselves. 
They can naturally be chosen as the operators that make up the quantum spin liquid Hamiltonian (e.g. see Eq.~\ref{eq:2dmodelapintro} and related discussion).
Next, we construct a set of mutually commuting operators by multiplying the check operators along the cardinal directions of the underlying lattice. 
These form the code stabilizer operators. 
They must commute with all the check operators and should be expressible as products of local check operators. 
Finally, we find non-local conserved operators winding around the lattice that encode the spin liquid's topological order. 
These must commute with all the check operators and the code stabilizers. 
Within this set, we identify pairs of non-commuting operators and choose them as the logical qubit operators of the topological subsystem code. 

Since all the code ingredients come from a topological quantum spin liquid, the topological order is \emph{built in} with the number of logical qubits (non-local conserved operators) determined by the lattice topology along with an inherent flexibility in regards to the lattice shape. 
The anticommuting local conserved operators determine the code's check operators and stabilizers, encoding a gauge redundancy and completing our construction of a topological subsystem code. 
In the sections that follow, we demonstrate this idea with two concrete examples on a square lattice and a kagome lattice. 
We will also show the error detection and correction procedure in these two examples. 
All of the above can be put on other lattices and graphs in a fairly straightforward manner.

\section{A weight-4 topological subsystem code on the square lattice}
\label{sec:weight4}

\begin{figure}
\centering
    \includegraphics[width=0.9\columnwidth]{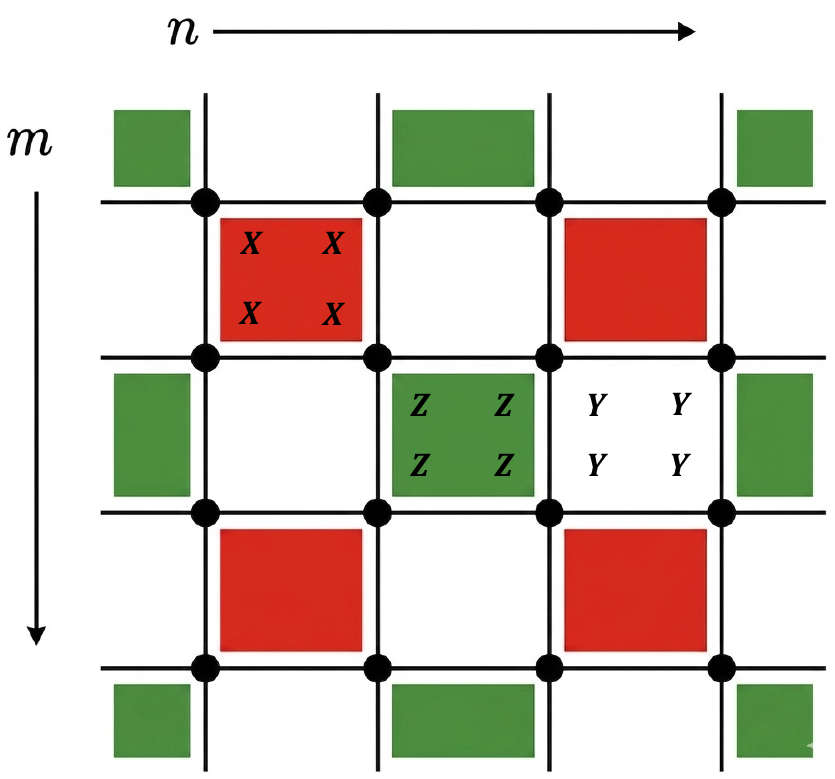} 
    \caption{Square lattice with qubits on the vertices. The $\boxed{x}$, $\boxed{z}$ and $\boxed{y}$ plaquette operators are shown by red, green and white colors respectively. The indices $\{m,n\}$ are row and column indices for the lattice. The $\boxed{x}$ and $\boxed{z}$ operators correspond to terms of the spin liquid Hamiltonian in Eq.~\ref{eq:2dmodelap}. They share a corner and mutually anticommute for nearest-neighbor plaquettes, in contrast to the mutually commuting arrangement of the toric code terms. In the associated error correction code here, the $\boxed{x},\boxed{z},\boxed{y}$ operators correspond to code check operators. }
    \label{fig:latticecheckops}
\end{figure}

Let us take the anticommuting quantum spin liquid model of Eq.~\ref{eq:2dmodelapintro} defined on the square lattice.
We write it down again below in Eq.~\ref{eq:2dmodelap} for convenience.
This will be used to construct a topological subsystem code with weight-4 check operators in what follows. 
We will explicitly state the code's check operators (gauge operators), stabilizer generators, logical qubit operators and enumerate the subsystem degree of freedom in the code. 
We will also discuss detection and correction of single qubit Pauli errors as well as Pauli string errors. 
Throughout this section, we consider a square lattice of linear dimension $L$ with periodic boundary conditions as shown in Fig.~\ref{fig:latticecheckops}. 
Each vertex of the lattice contains a qubit, giving us a total of $L^2$ qubits.
Here we have chosen an equal extent (of $L$) in both directions as convention which is not strictly necessary for our code constructions.

\subsection{Code check operators}
\label{sec:squarecheck}

The square lattice anticommuting spin liquid Hamiltonian 
\begin{equation}
    H =\; J_x \sum_{\boxed{x}} \left( \prod_{i \in\; \boxed{x}} 
    \sigma^x_i \right)
    +
    J_z \sum_{\boxed{z}} \left( \prod_{j \in\; \boxed{z}} 
    \sigma^z_j \right)
    \label{eq:2dmodelap}.
\end{equation}
has 4-spin plaquette operators that are shown in red and green colors in Fig.~\ref{fig:latticecheckops}. 
This model admits a set of local conserved operators that commute with the Hamiltonian in Eq.~\ref{eq:2dmodelap}, given by $\sigma^x_i \sigma^x_j \sigma^x_k \sigma^x_l$ on the $\boxed{z}$ plaquettes, $\sigma^z_i \sigma^z_j \sigma^z_k \sigma^z_l$ on the $\boxed{x}$ plaquettes and $\sigma^y_i \sigma^y_j \sigma^y_k \sigma^y_l$ on the $\boxed{y}$ plaquettes indicated by white plaquettes in Fig.~\ref{fig:latticecheckops}. 
Since the neighboring $\boxed{x}$ and $\boxed{z}$ plaquettes share a corner, the local conserved operators from neighboring $\boxed{x}$ and $\boxed{z}$ plaquettes do not commute but instead anticommute. 
As argued in Refs.~\cite{GSentropy_preprint,anticommutingqsl}, this anticommutation leads to an extensively degenerate quantum spin liquid ground state with topological order. 

We construct an associated topological subsystem code by choosing check operators as the minimal weight operators that commute with all the local conserved quantities in the quantum spin liquid. 
A natural choice therefore for these operators is the product of Pauli Z operators on the $\boxed{z}$ plaquettes, Pauli X operators on the $\boxed{x}$ plaquettes and Pauli Y operators on the  $\boxed{y}$ plaquettes. 
These are shown in Fig.~\ref{fig:latticecheckops} and can be formally written as
\begin{equation}\label{eq:checkops}
    C_\alpha = \prod_{i \in \boxed{\alpha}} \sigma^\alpha_i
\end{equation}
Note that the operators $C_X$ and $C_Z$ are same as the terms in the spin liquid Hamiltonian in Eq.~\ref{eq:2dmodelap}. 
The $C_Y$ operators are additional weight-4 operators that are linearly independent and preserve the conserved quantities of the quantum spin liquid. 
The $C_Y$ operators are going to be equally important for constructing the code space and correcting logical qubit errors.

\begin{figure*}
\centering
    \includegraphics[width=0.8\linewidth]{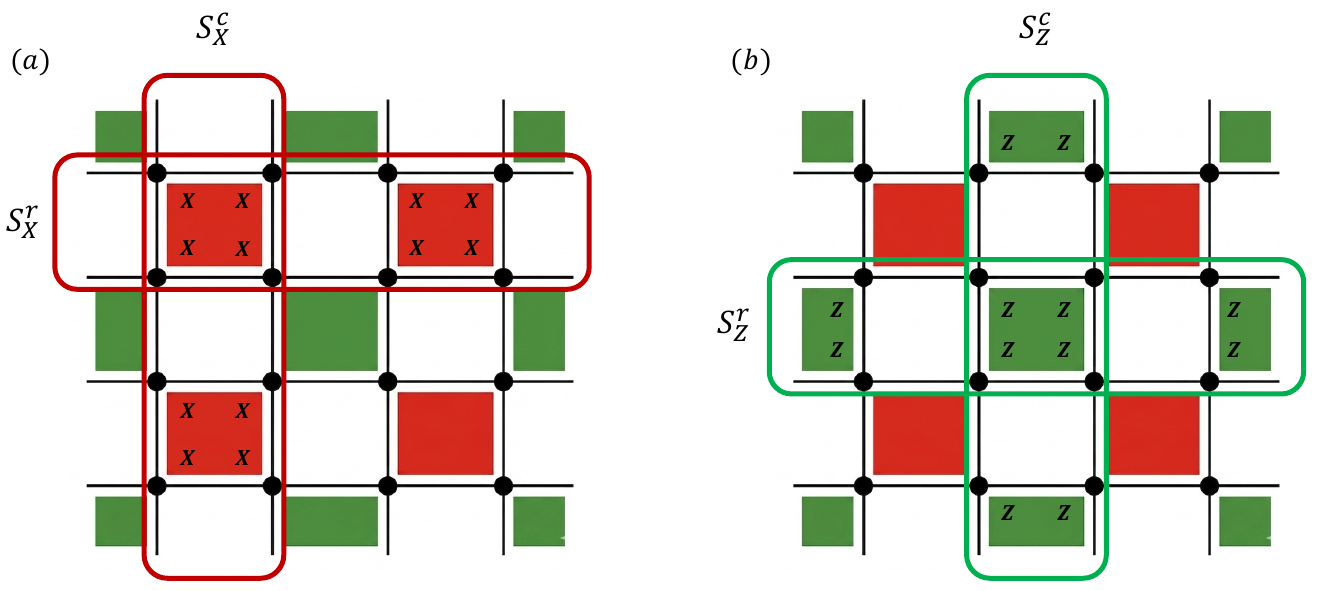}
    \caption{This figure schematically shows the code stabilizer operators of the square lattice ACC code. (a) shows examples of $S_X^r$ (Eq.~\ref{eq:SrX}) and $S_X^c$ (Eq.~\ref{eq:ScX}) indicated by the horizontal and vertical red rectangles, (b) shows $S_Z^r$ (Eq.~\ref{eq:SrZ}) and $S_Z^c$ (Eq.~\ref{eq:ScZ}) indicated by the horizontal and vertical green rectangles.}
    \label{fig:latticestabops}
\end{figure*}

Analogous to the check operators in the Bacon-Shor code, the check operators on $\boxed{x}$ and $\boxed{z}$ plaquettes that share a corner do not all mutually commute. 
These operators along with the set of anticommuting local conserved operators encode the gauge qubits, lending the code its subsystem degree of freedom. 
In contrast, the check operators on $\boxed{y}$ plaquettes mutually commute with each other and also commute with all the check operators on $\boxed{x}$ and $\boxed{z}$ plaquettes. 
Instead of the Bacon-Shor code, they are analogous to the mutually commuting weight-4 stabilizer operators in Kitaev's toric code. 

\subsection{Code stabilizers}\label{sec:squarelatticestabs}

After choosing the check operators, we can construct the set of code stabilizers. We require the stabilizers to be generated by the local check operators. Product of Pauli X and Pauli Z operators on $\boxed{x}$ and $\boxed{z}$ plaquettes respectively along a row (r) or column (c) are such examples of code stabilizer operators. The X code stabilizers along a row are 
\begin{equation}
    S^r_X = \prod_{j=1}^{L} \sigma^x_{2i-1,j}\sigma^x_{2i,j}
    \label{eq:SrX}
\end{equation}
Here $j$ is a column index and $i$ is an integer that goes from $1$ to $L/2$ such that $2i-1$ and $2i$ correspond to row indices. 
The row indices correspond to the \boxed{x} plaquettes that are present on alternating rows in Fig.~\ref{fig:latticecheckops}. 
Similarly, the other code stabilizers can be written as,
\begin{align}
     S^c_X = \prod_{j=1}^{L} \sigma^x_{j,2i-1}\sigma^x_{j,2i} \label{eq:ScX} \\
     S^r_Z = \prod_{j=1}^{L} \sigma^z_{2i,j}\sigma^z_{2i+1,j} \label{eq:SrZ} \\
     S^c_Z = \prod_{j=1}^{L} \sigma^z_{j,2i}\sigma^z_{j,2i+1} \label{eq:ScZ}
\end{align}
Note that in presence of periodic boundary conditions, $m=2i + 1$ with $i=L/2$ corresponds to $m = 1$. 
All operators in $S^r_X$, $S^c_X$, $S^r_Z$ and $S^c_Z$ are non-local and denoted in Fig.~\ref{fig:latticestabops}. 
However they can be generated by multiplying the local weight-4 check operators in $C_X$ and $C_Z$. 
According to our definition of code stabilizer operators, the check operators in $C_Y$ are also code stabilizers as they mutually commute with each other and also commute with all the other check operators. These code stabilizers are local in nature.

We now count the total number of stabilizer operators. There are $L/2$ independent stabilizers each in the sets $S^r_X$ and $S^c_X$. This implies there are $L$ such stabilizers. However if we multiply all the stabilizers present in the sets $S^r_X$ and $S^c_X$, we get the product of identity on all the qubits. Thus one of the stabilizers is not a linearly independent stabilizer, leaving $(L-1)$ linearly independent stabilizers across both these sets. By a similar argument, the number of independent stabilizer generators in the sets $S^r_Z$ and $S^c_Z$ are also $(L-1)$. For the set $C_Y$, there are a total of $L^2/2$ white plaquettes. However a product of all the stabilizers in $C_Y$ also gives a product of identity on all qubits. Thus there are only $L^2/2-1$ independent operators in $C_Y$.

Thus, in total, there are $2(L-1) + L^2/2-1$ code stabilizers. The code space is defined as the simultaneous eigenspace of all these code stabilizer operators. Traditionally, the chosen code space is the one where all code stabilizer operators are in their $+1$ eigenstate.

\subsection{Logical operators}

\begin{figure}
\centering
    \includegraphics[width=0.8\columnwidth]{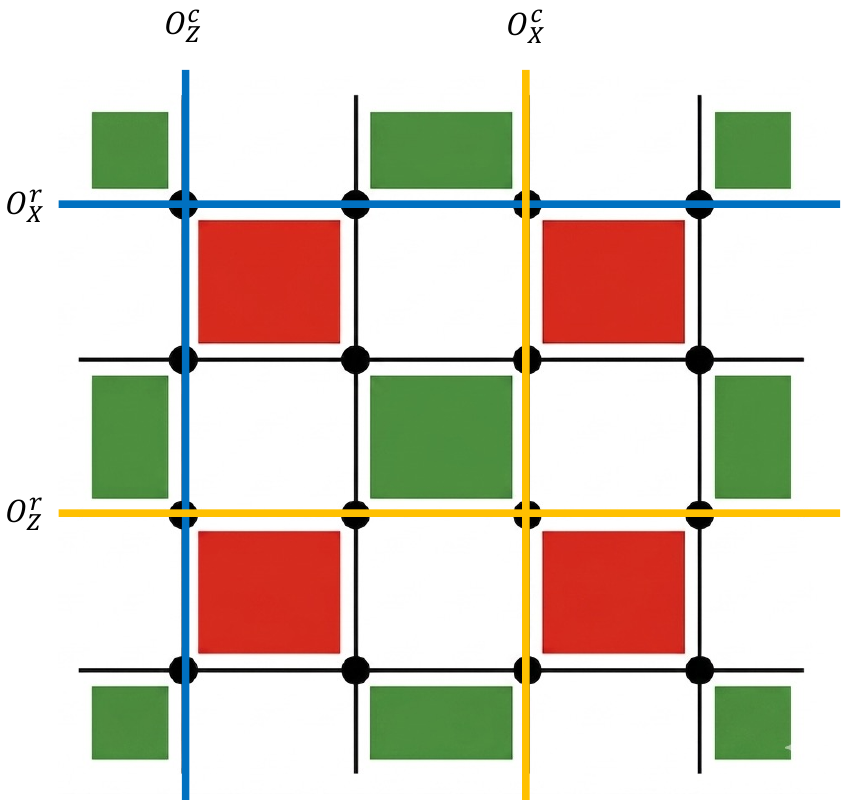} 
    \caption{This figure shows examples of encoded logical qubit operators for the square lattice ACC code. (a) Solid blue lines show one set of logical `X' and `Z' qubit operators as $O^r_X$ and $O^c_Z$. (b) Solid yellow lines show the other logical `X' and `Z' qubit operators as $O^r_Z$ and $O^c_X$. See Eqs.~\ref{eq:Orbeta},~\ref{eq:Ocbeta} for the definition of these operators.}
    \label{fig:latticelogicalops}
\end{figure}
Now we will construct the logical qubit operators for our square lattice code. 
We recall that a set of logical operators for an encoded logical qubit are chosen such that they mutually anticommute but commute with all the check operators and the code stabilizers. 
The topological order of the associated quantum spin liquid serves as the inspiration to construct these logical operators. 
Fig.~\ref{fig:latticelogicalops} shows the two logical X operators ($O^r_X,O^c_X$) and the two logical Z operators ($O^r_Z,O^c_Z$). These can be written as
\begin{align}
    O^r_\beta = \prod_{j=1}^{L} \sigma^\beta_{i,j} \label{eq:Orbeta} \\
    O^c_\beta = \prod_{j=1}^{L} \sigma^\beta_{j,i} \label{eq:Ocbeta}
\end{align}
for some (fixed) $i$ between $1$ and $L$ and $\beta \in \{x,z\}$.
Operators with different choices of the index $i$ are all equivalent upto multiplication by code stabilizer operators. 
The two sets $\{O^r_X, O^c_Z\}$ and $\{O^r_Z, O^c_X\}$ define the two logical qubits in our topological subsystem code. 

\subsection{The subsystem and topological properties of the square lattice code}

Our code space has a large subsystem degree of freedom due to the anticommuting check operators in $C_X$ and $C_Z$ vis-\`a-vis the anticommuting local conserved operators of the associated spin liquid. 
We enumerate this here. 
Based on our argument above in Sec.~\ref{sec:squarelatticestabs}, we can see that the code stabilizers and the two logical qubits fix $L^2/2 + 2L - 1$ qubit degrees of freedoms in any instance of the code. 
Out of $L^2$ qubits, this leaves the remaining $n = L^2/2 - 2L + 1$ degrees of freedom unspecified. These are the subsystem degrees of freedom. 
The code space has a Hilbert space dimension of $2^n$, i.e. there are $n$ gauge qubits. 
Transitions among states within this space is caused by the anticommuting check operators in $C_X$ and $C_Z$. 
Measurement of these check operators moves us within this $2^n$ dimensional code space but leaves the logical qubit and the code stabilizers unaffected. 
We also can see now that a subset of the gauge qubits may be chosen such that they are left unaffected even by check operator measurements. 
They correspond directly to the set of local conserved anticommuting operators as discussed in Sec.~\ref{sec:squarecheck}.

Analogous to the toric code, our code is also a topological code. 
In the toric code, the topological order stems from the dependence of the number of logical qubits on the topology of the lattice. 
This is true in our code as well. 
With periodic boundary conditions along both directions of the square lattice, there are two logical qubit operators that wrap around the lattice. 
Removing periodic boundary conditions along one of the directions destroys the corresponding logical qubit. 
Note that in the familiar Bacon-Shor subsystem error-correcting code, there is only one logical qubit regardless of the presence of periodic boundary conditions. 
Because of the topological nature of our code and the square lattice geometry, many of the techniques developed for the toric and surface codes~\cite{qecreview} should be applicable here as well in terms of physical implementation. 

\subsection{Single qubit error detection and correction}

Our code is able to detect and correct any single qubit Pauli error that takes us out of the code space. 
The procedure involves measuring the set of code stabilizers ($S_X^r,S_X^c,S_Z^r,S_Z^c$ and $C_Y$) as syndromes and applying a correction to bring us back into the code space. 
Due to the subsystem nature of our code, the non-local code stabilizers are constructed by \emph{sequentially} measuring the corresponding check operators (cf. Sec.~\ref{sec:squarelatticestabs}). 

\begin{figure}
\centering
    \includegraphics[width=\columnwidth]{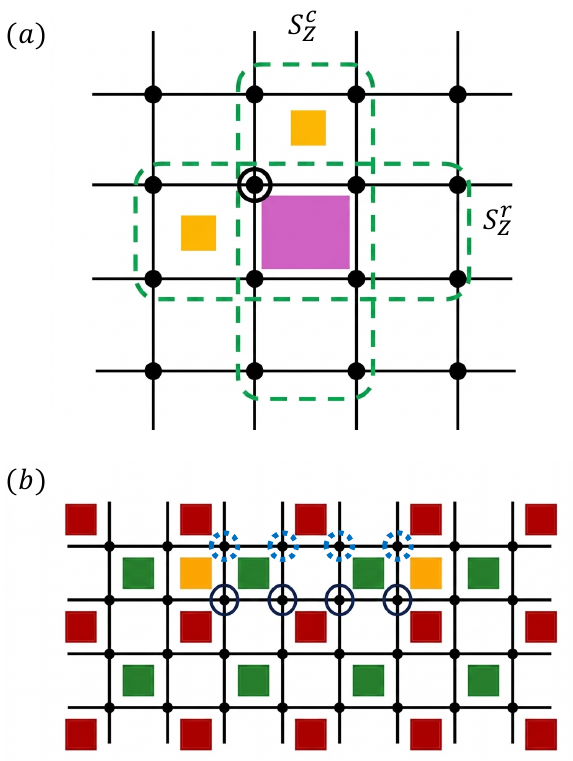} 
    \caption{(a) A single qubit Pauli X error shown by the circled vertex. The stabilizer error syndrome occurs in the dashed green boxes (involving a  $S^r_Z$ and a $S^c_Z$ stabilizer (Eq.~\ref{eq:SrZ},~\ref{eq:ScZ})) and the orange plaquettes (involving a $C_Y$ stabilizer (Eq.~\ref{eq:checkops}). (b) The vertices with black solid circles show a string of Pauli X errors with even number of sites. The orange boxes show the plaquettes in $C_Y$ that show an error syndrome. The vertices with blue dotted circles denote qubits that may be used to correct the errors.}
    \label{fig:errorcorrection}
\end{figure}

We first measure all the check operators in $C_X$. 
Their products along a row and a column gives us the measurement of stabilizers in the sets $S^r_X$ and $S^c_X$ respectively. 
Similarly, we then measure all the check operators in $C_Z$ to get measurement of the stabilizers in $S_Z^c$ and $S_Z^r$ sets. 
Since the check operators in $C_X$ and $C_{Z}$ operators do not commute when sharing corners, we have to do these measurements sequentially. 
We also measure the code stabilizers in the set $C_Y$. 
The $C_Y$ measurements can in principle be done independently in parallel to the earlier $C_X$, $C_Z$ measurements as the $C_Y$ operators are stabilizer generators and commute with all the check operators.
The stabilizers which show a measured eigenvalue of $-1$ are the syndromes and they help us detect the location of the single qubit Pauli error. 

Fig.~\ref{fig:errorcorrection}(a) shows a single qubit Pauli $X$ error on the circled qubit. 
It would show up as a syndrome in one stabilizer each in $S_Z^r$ and $S_Z^c$, shown by dashed green lines. 
This localizes the error onto a single plaquette that is colored pink in Fig.~\ref{fig:errorcorrection}(a). 
In parallel, a syndrome error in two plaquette operators in the set $C_Y$ (colored orange) then point to the exact site location of the error. 
Applying a Pauli X operator on this qubit restores the system into the code space. 
Additionally, applying Pauli X operators on the rest of the three qubits on the pink plaquette also brings us back into the code space. 
This latter operation would effectively amount to applying a loop of Pauli X operators on the pink plaquette. 
Recall that this operator is one of the local anticommuting conserved operators. 
It forms part of the redundant subsystem of our code and thus it does not affect the code space.
Although the logical qubit is exactly restored through this error correction procedure, the system in general moves within the much larger $2^n$-dimensional code space due to the sequential measurement of the non-commuting check operators. 
Measurement results of the individual non-commuting check operators can randomly be $\pm 1$, moving the system within this code space. 
Analogous considerations apply to a single qubit Pauli Z error that would be detected through appropriate $S_X^r$, $S_X^c$ and $C_Y$ syndromes, with the site location again pinpointed through appropriate $C_Y$ syndromes. 
For a single qubit Pauli Y = ZX error, appropriate $S_Z^r$, $S_Z^c$,$S_X^r$ and $S_X^c$ syndromes suffice to locate the error.

\subsection{Detecting and correcting a string of Pauli errors}

A string of Pauli errors can be corrected by the standard Minimum Weight Perfect Matching Decoder~\cite{mwpm}. 
The main idea is that a string of Pauli errors often lead to cancellation of syndromes along the chain, and the syndromes only show up at the ends of the string. 
Such is the case here as well and, given the endpoints, we can infer the shortest possible string and perform the error correction procedure based on it. 
This also puts a limit on the code distance, that is, the longest string of errors that can be reliably corrected. 

Our lattice is an $L \times L$ square lattice where the logical qubit operators are length $L$ operators that wrap around the lattice. 
An error string longer than $L/2$ can be incorrectly inferred to be the complementary shorter string that wraps around the lattice. 
Applying the correction on this shorter string would combine with the longer string error to generate a logical qubit operator, leading to an error in the logical qubit. 
This is quite analogous to e.g. the toric code or the Bacon-Shor code. 
Thus the code distance in our code is $L$. 
This makes our code an $[L^2,2,L]$ code where $L^2$ is the number of physical qubits, ``$2$'' denotes the number of encoded logical qubits and $L$ is the code distance. 
Note that the physical qubit count of this square lattice ACC code is half compared to the toric code and one-third compared to the subsystem surface code~\cite{toposubsys}.

As a concrete example, Fig.~\ref{fig:errorcorrection}(b) shows a horizontal Pauli X string error with even number of sites (shown by circled vertices). 
A syndrome of $-1$  only shows up in two of the stabilizers in the $C_Y$ set (colored orange in Fig.~\ref{fig:errorcorrection}(b)). 
To get the system back into the code space, we can apply Pauli X operators either on the solid black circled sites or the dotted blue circled sites. The former exactly cancels the error string while the latter is equivalent to applying plaquettes of Pauli X operators that are conserved and commute with the logical operators and the code stabilizers, leaving the code space unaffected. 

An analogous argument works for vertical string errors or strings of Pauli Z errors. 
Strings with odd number of sites can be corrected by combining the error correction of single qubits and even string length errors.
Recall that single qubit errors affect the $S_\beta^c$, $S_\beta^r$ sets, which will happen here as well and can be handled very similar to what has been described in the previous subsection.
Syndrome from Pauli Y errors are inferred by assuming that both Pauli X and Z errors have happened. 
Strings containing a mixture of Pauli X, Y and Z errors can be reconstructed by
multiplying strings that contain purely X or Z errors. 

\section{A weight-3 topological subsystem code on the kagome lattice}
\label{sec:weight3}

\begin{figure}
\centering
    \includegraphics[width=\columnwidth]{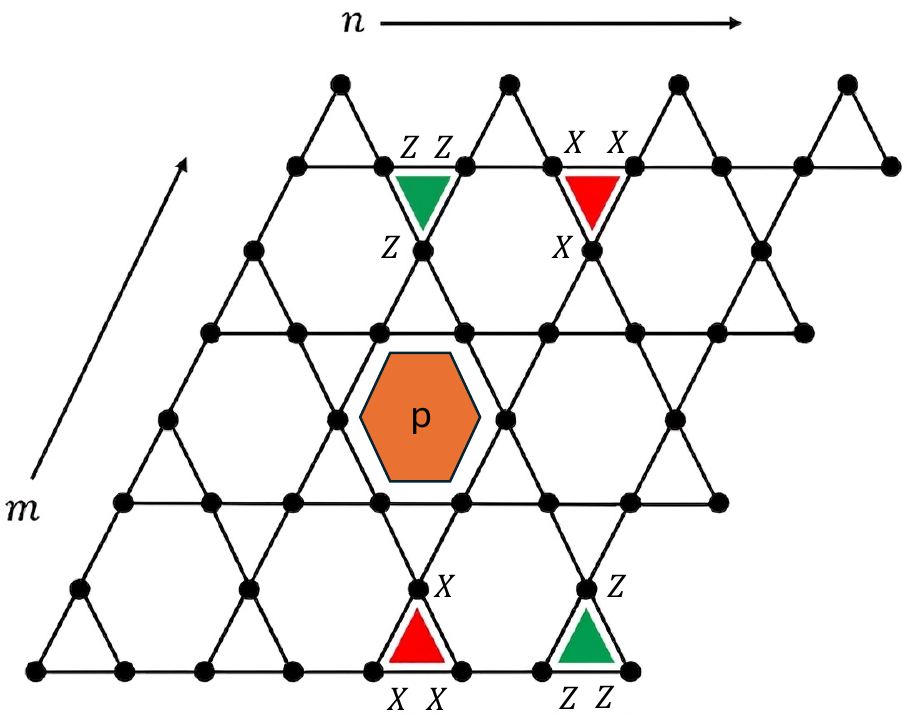} 
    \caption{Kagome lattice of size $L \times L$ with qubits on the vertices. $L=4$ in this figure. The indices $\{m,n\}$ are row and ``column'' indices for the lattice, ranging from $1$ to $2L$. By column, we mean the slanted vertical direction indicated by the arrow for the $m$ index throughout this section. The odd (even) index rows and columns contain $2L$ ($L$) vertices. The check operators are weight-3 operators, XXX and ZZZ on the triangular plaquettes shown in red and green colors respectively. The hexagonal plaquettes (such as the orange plaquette $p$) support local anticommuting conserved operators described by Eq.~\ref{eq:kagomeconserved}.}
    \label{fig:kagomecheckops}
\end{figure}

In this section, we convert the anticommuting quantum spin liquid model defined on a kagome lattice into a topological subsystem code with weight-3 check operators. 
Just like our square lattice example, we will explicitly construct the code stabilizers, logical operators, and error detection and correction procedure. 
This would show the inherent geometrical flexibility of our code construction template. 
Throughout this section, we choose a kagome lattice of linear dimension $L$ as shown in Fig.~\ref{fig:kagomecheckops} with periodic boundary conditions. 
Each vertex contains a qubit, giving us a total of $3L^2$ physical qubits.

\begin{figure*}
\centering
    \includegraphics[width=14cm]{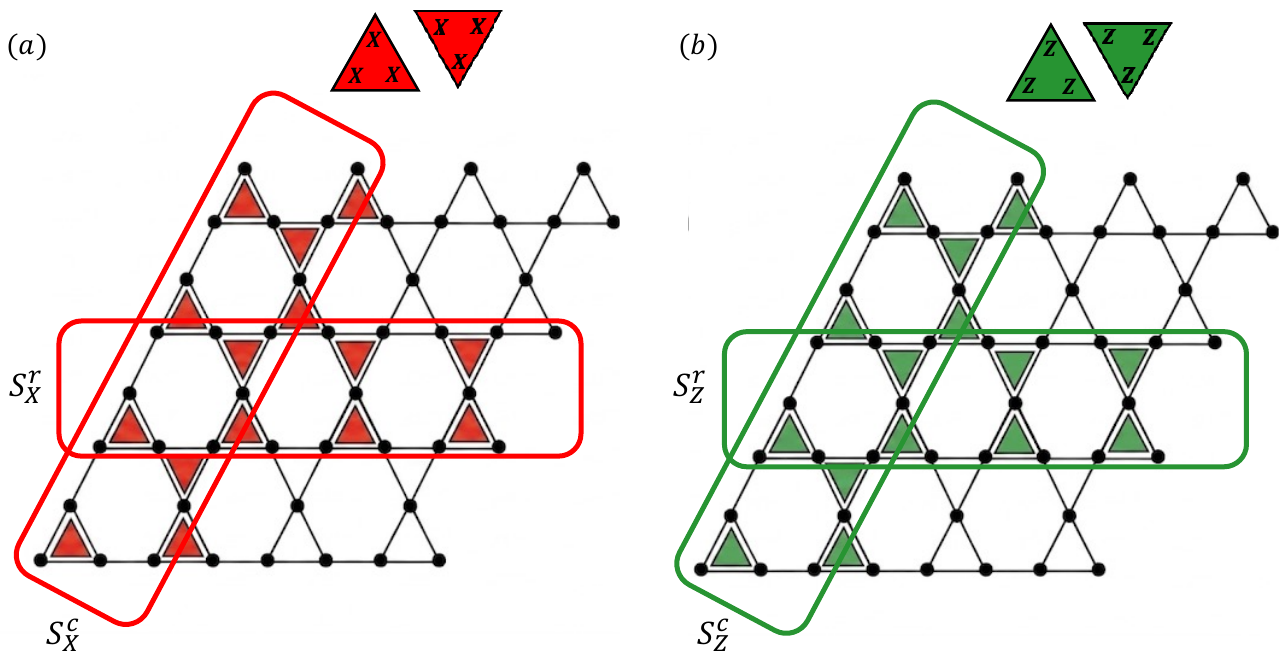}
    \caption{
    This figure schematically shows the code stabilizer operators of the kagome lattice ACC code. (a) shows examples of $S_X^r$ and $S_X^c$  indicated by the horizontal and slanted vertical red rectangles. (b) shows $S_Z^r$  and $S_Z^c$ indicated by the horizontal and slanted vertical green rectangles. 
    See Eqs.~\ref{eq:Srkagome},~\ref{eq:Sckagome} for definitions.
    Note that each horizontal (slanted vertical) rectangle appears to include three rows (columns) of qubits, however the middle row (column) gets excluded from the stabilizer operator. 
    This is because these stabilizers are generated as products of check operators on triangles, which leads to the identity operator on all the qubits in the middle row (column).} 
    \label{fig:kagomestabops}
\end{figure*}

\subsection{Code check operators}

The anticommuting spin liquid Hamiltonian defined on the kagome lattice~\cite{anticommutingqsl} is given by
\begin{equation}
    H =\; J_x \sum_{\Delta} \left( \prod_{i \in\; \Delta} 
    \sigma^x_i \right)
    +
    J_z \sum_{\nabla} \left( \prod_{j \in\; \nabla} 
    \sigma^z_j \right)
    \label{eq:2dmodelkagome}.
\end{equation}
Here $\Delta$ and $\nabla$ denote the two types of triangular plaquettes in the kagome lattice in Fig.~\ref{fig:kagomecheckops}. 
This model admits a set of local conserved operators given by
\begin{equation}\label{eq:kagomeconserved}
    D_\alpha = \prod_{i \in p} \sigma^\alpha_i.
\end{equation}
Here $p$ are the hexagonal plaquettes (example shown in Fig.~\ref{fig:kagomecheckops} as an orange plaquette), $i$ is a site index on $p$ and $\alpha \in \{x,z\}$. 
Since the neighboring hexagonal plaquettes share a corner, these local conserved operators do not all mutually commute. 
Similar to the square lattice, Refs.~\cite{GSentropy_preprint,anticommutingqsl} show that this anticommutation leads to an extensively degenerate quantum spin liquid ground state with topological order. 

To construct the associated topological subsystem code, a natural choice for the check operators is the product of Pauli X and Pauli Z operators on all the triangular plaquettes of the lattice. 
These check operators thus may be written as
\begin{equation}
    C_\alpha = \prod_{i \in t} \sigma^\alpha_i
    \label{eq:kagome_checkops}
\end{equation}
where $\alpha \in \{x,z\}$ and $i$ is a site index on a triangular plaquette denoted by $t$. 
These are weight-3 check operators that commute with all the local conserved operators defined in Eq.~\ref{eq:kagomeconserved}.
They do not all mutually commute with each other. 
Thus this set of check operators and the anticommuting local conserved operators on the hexagonal plaquettes form the gauge qubits, leading to the subsystem degree of freedom.

\subsection{Code stabilizers}

Having chosen the check operators, we can identify the code stabilizers as the operators that are generated by the local check operators and that mutually commute with each other. 
Products of Pauli X and Z operators on triangular plaquettes along two consecutive rows and columns are examples of such operators. 
They are shown in Fig.~\ref{fig:kagomestabops}. 
More formally, we can write them as
\begin{equation}
    S^r_\alpha = \prod_{j=1}^{2L} \sigma^\alpha_{2i-1,j}\sigma^\alpha_{2i+1,j} \label{eq:Srkagome}
\end{equation}
and
\begin{equation}
    S^c_\alpha = \prod_{i=1}^{2L} \sigma^\alpha_{i,2j-1}\sigma^\alpha_{i,2j+1}. \label{eq:Sckagome}
\end{equation}
Here $i$ and $j$ are row and column indices on the kagome lattice with periodic boundary conditions. 
Examples of stabilizers $S^r_\alpha$ ($S^c_\alpha$) are shown as horizontal (slanted vertical) red and green rectangles in Fig.~\ref{fig:kagomestabops}.
These code stabilizers are non-local but are generated by local weight-3 check operators on triangular plaquettes. We can now enumerate the number of code stabilizers. 
There are $(L-1)$ independent stabilizers each in the sets $S^c_{x/z}$ and $S^r_{x/z}$, giving us a total of $4(L-1)$ code stabilizers.
The code space is the simultaneous eigenstate of these code stabilizers which is conventionally chosen to be their $+1$ eigenstate.  

\subsection{Logical Operators}

\begin{figure}
\centering
    \includegraphics[width=\columnwidth]{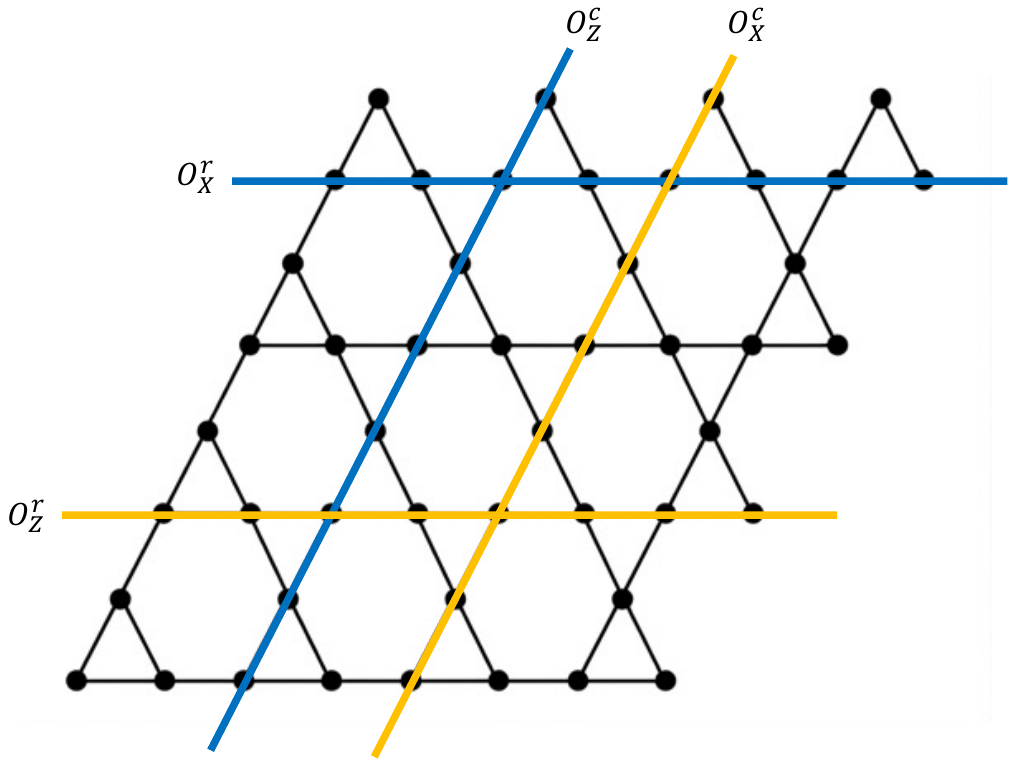} 
    \caption{
    This figure shows examples of encoded logical qubit operators for the kagome lattice ACC code. (a) Solid blue lines show one set of logical `X' and `Z' qubit operators as $O^r_X$ and $O^c_Z$. (b) Solid yellow lines show the other logical `X' and `Z' qubit operators as $O^r_Z$ and $O^c_X$. See Eqs.~\ref{eq:Orkagome},~\ref{eq:Ockagome} for the definition of these operators. }
    \label{fig:kagomelogicalops}
\end{figure}

\begin{figure*}
\centering
    \includegraphics[width=14cm]{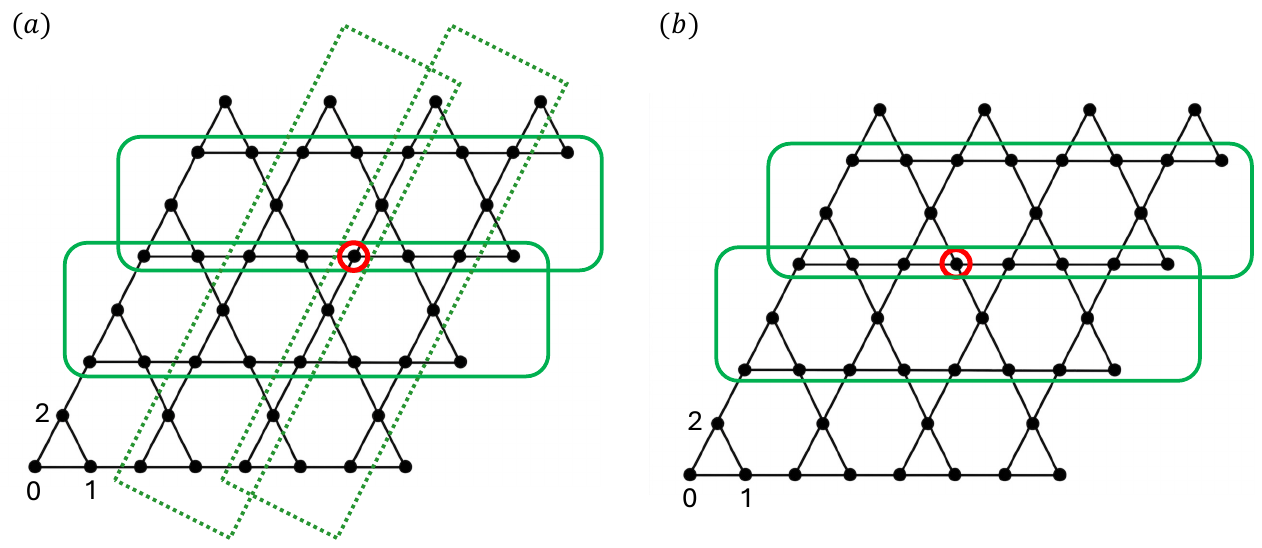} 
    \caption{The numbers $\{0,1,2\}$ mark the three distinct site types on the kagome lattice. Each site type corresponds to a unique error correction procedure. A single qubit Pauli error denoted by a qubit circled red on (a) site type '0' and (b) site type '1'. The solid and dotted rectangles denote the $S^r_Z$ and $S^c_Z$ stabilizers (Eqs.~\ref{eq:Srkagome},~\ref{eq:Sckagome}) that show the error syndrome in the two scenarios.}
    \label{fig:kagomeerror}
\end{figure*}

Armed with the check operators and code stabilizers, we now identify a set of logical qubit operators that mutually anticommute but commute with all the check operators and code stabilizers. 
Analogous to the square lattice example that we saw before, we can encode two logical qubits with logical X operators $\{O^r_X,O^c_X\}$ and logical Z operators $\{O^r_Z,O^c_Z\}$ owing to the many-body topological order of the associated anticommuting quantum spin liquid. 
These are shown as the blue and orange lines in Fig.~\ref{fig:kagomelogicalops} and written as
\begin{align}
    O^r_\alpha = \prod_{j=1}^{2L}\sigma^\alpha_{2i-1,j} \label{eq:Orkagome} \\
    O^c_\alpha = \prod_{i=1}^{2L}\sigma^\alpha_{i,2j-1} \label{eq:Ockagome}
\end{align}
where $\alpha \in \{x,z\}$. 
The two sets, $\{O^r_X,O^c_Z\}$ and $\{O^c_X,O^r_Z\}$ intersect exactly at one site. 
Thus they anticommute and define the two encoded logical qubits in our code.


\subsection{Code's subsystem and topological properties}

Similar to the square lattice case, this code here also has a large subsystem degree of freedom. 
The kagome lattice of size $L \times L$ has $3L^2$ qubit degrees of freedom. 
We saw earlier that the code has $4(L-1)$ code stabilizers and 2 logical qubits.
These together specify only $(4L-2)$ qubit degrees of freedom, leaving the remaining $ n = (3L^2 - 4L + 2)$ qubit degrees of freedom unspecified which are gauge qubits. 
Note that not all the gauge qubits are affected by measuring the check operators. 
These correspond to the local anticommuting conserved operators defined in Eq.~\ref{eq:kagomeconserved} similar to the square lattice case.

The number of encoded logical qubits in our code is dependent on the topology of the lattice. 
With periodic boundary conditions along both the cardinal directions of the lattice, there are two logical qubits as we saw earlier. 
Removing periodic boundary conditions along a direction introduces dangling bonds at the edges that do not form a complete triangular plaquette. 
Operators defined on these bonds do not commute with the corresponding logical operators, destroying its encoding. 
This dependence on the lattice topology and the fact that the code stabilizers are generated by local check operators makes our code a topological code.

\subsection{Error detection and correction}

The kagome code can detect and correct for errors quite analogously to the square lattice code.
Errors are detected by measuring the code stabilizers and identifying which stabilizers are no longer in their $+1$ eigenstate, specifying the syndrome. 
Based on that, we apply a correction that restores all the stabilizers back to the code space, thereby recovering the logical qubit(s). 
The code stabilizers in $S^r_X,S^r_Z,S^c_X$ and $S^c_Z$ are measured by sequentially measuring all the local check operators in $C_X$ and $C_Z$. 
Multiplying the measured values of the check operators along a row or column gives us access to the eigenvalues of the code stabilizers. 

In Fig.~\ref{fig:kagomelogicalops}, we have shown our kagome lattice with three distinct site types in terms of error locations labeled by 0, 1 and 2. 
The three distinct site types arise due to the structure of our code stabilizers and logical operators. 
Site type 1 does not occur within the stabilizers $S^c_\alpha$ and the logical operators $O^c_\alpha$. 
Similarly, site type 2 does not appear in the stabilizers $S^r_\alpha$ and the logical operators $O^r_\alpha$. 
Site type 0 on the other hand is involved in all the code stabilizers and logical operators.
Starting with single qubit Pauli errors, there is a distinct error correction procedure for each of the site types.
First, let's consider a Pauli X error on site of type 0, as shown in Fig.~\ref{fig:kagomeerror}(a) with a red circle around the error qubit. 
This would show up as a syndrome in two code stabilizers each in $S^c_Z$ and $S^r_Z$ as shown by green dotted and solid rectangles in Fig.~\ref{fig:kagomeerror}(a). 
The intersection of these exactly denotes the location of the single qubit error. 
Applying a Pauli X operation on this qubit then restores us back into the code space. An analogous argument works for Pauli Z errors.

Next we consider a Pauli X error on site type 1 shown as a red circle in Fig.~\ref{fig:kagomeerror}(b). 
This would show up as a syndrome in two stabilizers each in $S^r_Z$ as shown by solid green rectangles in Fig.~\ref{fig:kagomeerror}(b). 
The two stabilizers intersect in a line and thus the error qubit is only localized upto qubits of site type 1 along this line. 
Applying a Pauli X operator on any of these qubits restores the system back into the code space. 
This is because a two-qubit Pauli XX operator on qubits of site type 1 along a line commutes with the code stabilizers and the logical qubit operators. 
An analogous argument works for Pauli X errors on site type 2. 
Replacing X by Z in the above discussion specifies the error correction procedure due to a Pauli Z error as well.

A Pauli string error can be corrected using the Minimum Weight Perfect Matching Decoder~\cite{mwpm}, as strings connecting different sites of the same site type cancel out syndromes within the string. Only the end points of the strings are detected through syndromes. 
Similar to the square lattice argument, this limits the code distance to $L$ on our $L \times L$ kagome lattice. 
Thus our code becomes a $[3L^2,2,L]$ code. In contrast to the square lattice, the kagome lattice code requires three times as many qubits for the same number of encoded logical qubits. 
However kagome lattice has the lower weight-3 check operators. 
The kagome lattice ACC code has the same parameters as the subsystem surface code~\cite{toposubsys}.

Note that an even length string of Pauli X or Z errors on site type 1 or 2 does not cause an error as it commutes with all the code stabilizers and the logical operators as already encountered when discussing single qubit errors. 
Thus errors on these site types would always show up as single qubit errors. 
An error string that simultaneously includes all site types can be corrected by inferring a string on site type 0 and a single qubit error on site type 1 or 2 followed by the decoder. 

\section{Conclusions}
\label{sec:conclusion}

In summary, we introduced a recipe or framework for constructing topological subsystem codes from anticommuting quantum spin liquid models proposed recently in Refs.~\cite{GSentropy_preprint,anticommutingqsl}. 
The multi-spin models from this class host many-body topological order along with a massive degeneracy due to the presence of an extensive number of conserved local anticommuting operators. 
These two properties form the basis of the topological and subsystem nature respectively of the resulting error-correcting codes. 
These codes can combine the benefits of both topological stabilizer codes (such as toric code) and subsystem codes (such as Bacon-Shor code) for fault-tolerant quantum computation on current quantum hardware. 
The presence of local conserved anticommuting operators within the code creates a new category of topological subsystem codes that we termed as anticommuting charge (ACC) codes. 
These charges can be thought of as a subset of the gauge qubit degrees of freedom that is left undisturbed while the code stabilizers are being measured through local check operators. 
In this work, this gauge qubit subset was simply in the background and did not actively participate in the error correction. 
It would be interesting to design protocols to store and process additional quantum information using these undisturbed gauge qubits and determine whether they can improve the threshold error rates in these codes.

\begin{figure}
    \includegraphics[width=0.8\linewidth]{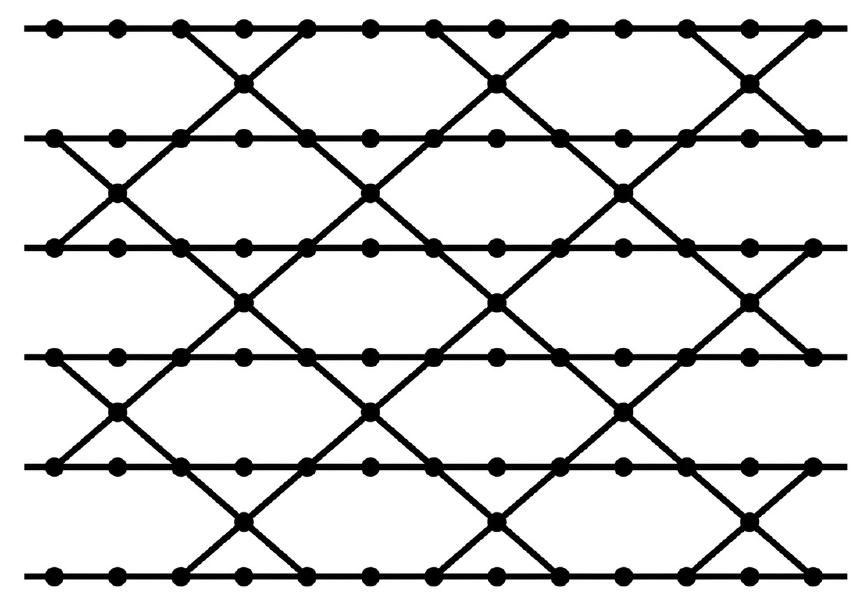}
    \caption{Kagome geometry (denoted by black solid lines) superimposed on IBM's Eagle~\cite{ibm,ibm_eagle} quantum processor's ``heavy-hex'' layout of qubits (denoted by black solid dots).}
    \label{fig:existing_schematic}
\end{figure}

Our template allows us to construct these topological subsystem codes on different lattice geometries, lending a flexibility that is not present in prior constructions. 
We show this explicitly by presenting detailed examples on a square lattice and a kagome lattice, requiring weight-4 and weight-3 check operator measurements respectively. 
Such a flexibility can be exploited for finding topological subsystem codes in a variety of other settings that may enhance encoding rates and threshold error rates. 
It will also be suitable for implementation on quantum hardware that might favor certain specific connectivities and geometrical layouts that may not be amenable to the earlier rigid constructions. 
As an example, we schematically show a kagome-like corner-sharing graph superimposed on the existing physical qubit layout of IBM's Eagle processor~\cite{ibm,ibm_eagle} in Fig.~\ref{fig:existing_schematic}.
Google's Willow chip~\cite{googlewillow,google_willow} and Rigetti's Cepheus processor~\cite{rigetti_cepheus} could serve as natural platforms for the square lattice ACC code.

The examples we present require at least weight-3 check operator measurements in contrast to the weight-2 operator construction by Bombin on trivalent lattices~\cite{Bombin_2010}. 
This leads to the natural question of whether our template allows topological subsystem codes with weight-2 operator measurements. 
There do exist Hamiltonian models consistent with the central idea of Hamiltonians having anticommuting terms sharing a corner/single site on the lattice. 
This includes Kitaev's honeycomb model~\cite{Kitaev_2006} and quantum-compass models~\cite{Nussinov_vandenBrink_review_2015,Brzezicki_Dziarmaga_Oles_2007,Dorier_Becca_Mila_2005,Jackeli_Khaliullin_PRL_2009,Mizoguchi_2022}. 
Kitaev's honeycomb model has topological order but encodes no logical qubits when seen as a subsystem code~\cite{SBT11}. 
The 1D compass model has Majorana edge modes~\cite{zeromodescompass} but does not possess a commuting set of local stabilizers for error correction and detection. 
The 2D quantum compass model leads to the Bacon-Shor subsystem code which is not topological and has a rigid geometry~\cite{Bombin_2010}. 
Coming up with models with either local or non-local weight-2 operators within our framework would be an interesting topic for the future.
Another natural quantitative issue would be to perform threshold computations in order to characterize the performance of these ACC codes to move towards potential application. 
For instance, gauge fixing in subsystem codes has previously been shown to improve error thresholds~\cite{higgott_breuckmann_prx2021}.

In terms of outlook, this identification of a class of topological subsystem codes based on anticommuting quantum spin liquids opens up new directions of research. 
For instance, it would be fruitful to explore the properties of excitations in these codes analogous to the $e$ and $m$ excitations in the toric code. 
Given the subsystem feature of these codes, one can ask as to their connections to fracton models possessing subsystem symmetries~\cite{fractons}. 
Some aspects of this were discussed in Ref.~\cite{anticommutingqsl}.
Another compelling direction would be to study non-local version of these codes on different graphs in order to encode a higher number of logical qubits in the spirit of good quantum Low Density Parity Check (qLDPC) codes~\cite{qldpc1,qldpc2,qldpc3,qldpc4,qldpc5}. 
There have been extensive studies of measurement-induced phases such as spin liquids and spin glasses from topological~\cite{measspinliquid,measspinliquid1} and subsystem codes~\cite{subsystemspinglass} respectively. 
It remains to be seen what measurement-induced phases and entanglement structure~\cite{entstruc} can be produced by combining topological order and subsystem redundancy in error-correcting codes such as the ones presented here.

\vspace{1cm}
\begin{acknowledgements}
Discussions with Beno\^it Dou\c cot and Pavithran Iyer are gratefully acknowledged. 
SP is also thankful to Amir Hamza Shaikh, Harsh Sharma and Himadri Sekhar Dhar for general quantum error correction discussions.
V.S. acknowledges support from the J. Evans Attwell Welch fellowship by the Rice Smalley-Curl Institute.
S.P. acknowledges funding support from SERB-DST, India (superseded by ANRF-DST established through an Act of Parliament: ANRF Act, 2023) via Grant No. MTR/2022/000386.
S.P. also acknowledges the Indo-Japan LOTUS science exchange award for partial support during the final stages of this project.
\end{acknowledgements}

\bibliographystyle{unsrt}
\bibliography{refs.bib}

\end{document}